# GAD: A Real-time Gait Anomaly Detection System with Online Adaptive Learning

Ming-Chang Lee[1], Jia-Chun Lin[2] and Sokratis Katsikas[3]

[1,2,3]*Department of Information Security and Communication Technology, Norwegian University of Science and Technology,*
*Gjøvik, Norway*

[1] *mingchang1109@gmail.com*
[2]*jia-chun.lin@ntnu.no*
[3] *sokratis.katsikas@ntnu.no*

5th May 2024

# GAD: A Real-time Gait Anomaly Detection System with Online Adaptive Learning

Ming-Chang Lee[1][0000−0003−2484−4366], Jia-Chun Lin[1][0000−0003−3374−8536], and Sokratis Katsikas[1][0000−0003−2966−9683]

Department of Information Security and Communication Technology, Norwegian University of Science and Technology (NTNU), Gjøvik, Norway
mingchang1109@gmail.com,{jia-chun.lin,sokratis.katsikas}@ntnu.no

**Abstract.** Gait anomaly detection is a task that involves detecting deviations from a person's normal gait pattern. These deviations can indicate health issues and medical conditions in the healthcare domain, or fraudulent impersonation and unauthorized identity access in the security domain. A number of gait anomaly detection approaches have been introduced, but many of them require offline data preprocessing, offline model learning, setting parameters, and so on, which might restrict their effectiveness and applicability in real-world scenarios. To address these issues, this paper introduces GAD, a real-time gait anomaly detection system. GAD focuses on detecting anomalies within an individual's three-dimensional accelerometer readings based on dimensionality reduction and Long Short-Term Memory (LSTM). Upon being launched, GAD begins collecting a gait segment from the user and training an anomaly detector to learn the user's walking pattern on the fly. If the subsequent model verification is successful, which involves validating the trained detector using the user's subsequent steps, the detector is employed to identify abnormalities in the user's subsequent gait readings at the user's request. The anomaly detector will be retained online to adapt to minor pattern changes and will undergo retraining as long as it cannot provide adequate prediction. We explored two methods for capturing users' gait segments: a personalized method tailored to each individual's step length, and a uniform method utilizing a fixed step length. Experimental results using an open-source gait dataset show that GAD achieves a higher detection accuracy ratio when combined with the personalized method.

## 1 Introduction

Human gait refers to the style of walking of an individual. It encompasses the motion and pattern of limbs and body during locomotion, influenced by many factors such as weight, limb length, footwear, health conditions, and other personal characteristics [9]. Human gait is recognized as one of biometric measures for identifying individuals due to its distinctive and complex pattern unique to each person. Researchers also analyzed gait for diagnosing, tracking, and evaluating treatments for various diseases and neurodegenerative disorders [1,12].

Detecting anomalies in gait data is crucial for identifying irregularities that may indicate health conditions, thereby improving overall well-being through timely healthcare interventions. In the security domain, gait anomaly detection also plays an important role as it enhances personal security by enabling biometric verification, thus preventing unauthorized access or impersonation.

A number of gait anomaly detection approaches have been introduced, and they can generally be classified into wearable methods and non-wearable methods [18]. The former requires users to wear devices equipped with sensors such as accelerometers or gyroscopes, offering direct and continuous monitoring but potentially intrusive, while the latter relies on external systems like video cameras or floor sensors, providing a more passive form of detection but often requiring specific environments and setups. Due to the fact that wearable sensors can be easily attached to various parts of the body without requiring a specific environment, they are ideal for collecting and monitoring human gait [7,17].

However, approaches based on wearable devices focus on extracting features and engineering optimal features [8]. For example, the authors in [2] utilized a Shimmer 2R sensor device to extract 11 acceleration-based features. Similarly, multiple features were extracted and generated in studies [8] and [16]. However, as noted in [8], estimating these features often requires additional event detection and professional expertise to effectively utilize the collected data. Furthermore, the manual extraction of features for machine learning-based systems is usually susceptible to bias, particularly due to the complex nature of sensor data.

Several deep learning-based approaches have been introduced to address the issues related to manual feature extraction, as they are capable of learning the pattern of gait directly from raw sensor data. Example solutions include [8,20]. However, many of the existing deep learning-based approaches require offline data preprocessing, offline model learning, and/or setting or tuning parameters. These requirements might lead to delays in anomaly detection and could limit their effectiveness in real-time applications.

In this paper, we introduce GAD, a real-time gait anomaly detection system based on dimensionality reduction and Long Short-Term Memory (LSTM), which is a type of recurrent neural networks. GAD focuses exclusively on three-dimensional accelerometer data, which includes readings along the x, y, and z axes. This type of data is available in various wearable devices, such as smartphones, smartwatches, and specialized sensor bands, thereby enhancing GAD's broad applicability. Unlike other methods that generate or rely on multiple features extracted from accelerometer data, GAD simplifies data processing by converting the three-dimensional accelerometer data into a single-variable series.

Upon initiation, GAD begins converting each accelerometer instance received from the user into a unified value to form a short gait segment. Concurrently, GAD trains an anomaly detector to learn the user's walking pattern within the gait segment. The detector is designed with two LSTM-based anomaly detection models that are trained in a partially parallel manner. Instead of training the models with the original gait segment, GAD converts the segment into a less complex AARE (Average Absolute Relative Error) series and uses this series to train the first detection model. Simultaneously, the AARE series is further converted into another AARE series (i.e., another dimensionality reduction), and the resulting AARE series is utilized to train the second detection model. Both models feature a simple and lightweight LSTM network structure (1 hidden layer with 10 hidden units), and they are trained and retrained based on their most recent input values in a sliding window manner.

Once the anomaly detector is generated, the user undergoes a model veri­fication process with their two subsequent steps. If the verification fails, GAD prompts the user to restart the entire process. Conversely, if the verification is successful, the anomaly detector will be utilized to detect anomalies in the user's future gait readings upon the user's request. In this case, the anomaly detector will continue to be retrained when it experiences minor pattern changes or when it cannot provide adequate predictions.

To evaluate the detection performance of GAD, we studied two methods for capturing users' gait segments. One is a personalized method that is tailored to each individual's step length. The other is a uniform method, which always utilizes a fixed step length for every individual. Our experiment results, using an open-source gait dataset, demonstrate that GAD delivers a satisfactory per­formance. It achieves a higher anomaly detection ratio when combined with the personalized method than when combined with the uniform method.

The rest of the paper is organized as follows: Section 2 describes related work. In Section 3, we detail the design of GAD. Section 4 presents and discusses the experiments conducted and their corresponding results. Finally, Section 5 concludes this paper and outlines our future work.

## 2    Related Work

The advancement in artificial intelligence has significant impact on various do­mains, including gait anomaly detection. Recent approaches to gait anomaly detection can be broadly classified into two categories: those based on machine learning and those based on deep learning.

The machine learning-based approaches typically involve traditional algo­rithms like decision trees, support vector machines, or k-nearest neighbors, which often require manual feature extraction and selection from gait data. For in­stances, Nukala et al. in [14] studied several classification algorithms, including the back propagation artificial neural network, support vector machine, k-nearest neighbors, etc., to classify patients and normal subjects based on features ex­tracted from the raw gait data collected from the gyroscopes and accelerome­ters. Otamendi et al. in [15] proposed a personalized, machine learning-based approach to detect significant changes in the functional state of individuals that require the use of Assistive Devices for Walking.

However, machine-learning based gait anomaly detection has drawbacks and limitations, including potential bias in the training data, the need for large datasets for effective training, the need for manual feature extraction or generation, consequently reducing their practicality in real-world applications [8].

Deep learning-based approaches, on the other hand, utilize neural networks, particularly convolutional and recurrent neural networks, to automatically learn and extract features directly from raw data. These approaches offer enhanced accuracy and the ability to capture more complex patterns in gait. Potluri et al. in [16] introduced a wearable sensor system designed to detect human gait abnormalities by integrating a plantar pressure measurement unit with Inertial Measurement Units (IMUs) and using a stacked Long Short-Term Memory model. However, the proposed approach has a limitation in wide applications due to the reliance on the plantar pressure measurement unit. Furthermore, it requires offline model training, which makes it less adaptable to dynamic changes in an individual's gait pattern.

Sadeghzadehyazdi et al. in [19] presented an end-to-end deep learning model that utilizes skeleton data from the Kinect to identify gait anomalies. To understand the relationship between various body joints during movement, the model captures spatial and temporal patterns by analyzing the entire skeleton. While the model showed promising results, the accuracy varied across different datasets, which indicates potential challenges in adapting to new gait patterns or populations. In addition, the model's reliance on Kinect for skeleton data might limit its applicability in real-world scenarios.

Cola et al. in [2] introduced a gait anomaly detection approach for continuous monitoring and detection of changes in gait patterns using a single wearable device equipped with a three-dimensional accelerometer. The approach extracts 11 acceleration-based features from 43 features using a greedy heuristic approach. A personalized training set is created with the gait segments recorded during the initial days of use, and this data is to train a binary k-nearest neighbors classifier. The strength of this method lies in its low complexity and computational requirements, allowing the algorithms to run directly on the wearable device for real-time analysis. However, as the authors noted in the paper, the detection performance of the approach relies on two parameters: the number of neighbors ($k$) and the coverage index ($c$). Additionally, the approach requires some time to collect training data, but it is unclear how much data would be sufficient. Contrary to the above-mentioned approach, the GAD proposed does not require gait data to be collected for several days or to extract additional features from the original accelerometer data. Furthermore, GAD is parameter-free, which removes the need for users to determine any specific settings.

## 3   Design of GAD

As previously stated, GAD focuses on three-dimensional accelerometer data, which is commonly available in various wearable devices and include readings along the x, y, and z axes. In this paper, we refer to each collected/observed accelerometer data point, consisting of three-dimensional readings, as an accelerometer instance. GAD consists of three main components:

– Base Model Generator (BaseGen): Aims to generate an anomaly detector for the target user.
– Model verification: Responsible for validating the created anomaly detector.
– Online Anomaly Detection (OLAD): Responsible for anomaly detection, as well as online model retraining and adaptation.

| | |
|---|---|
| **Input:** Requests and accelerometer instances observed from the user | |
| **Output:** Notification to user requests | |
| **Procedure:** | |

```
1:    if receiving a request for model generation {
2:        Let i = 1; // Initializes index for accelerometer instances;
3:        Let f_dect be false; //No anomaly detector created;
4:        Let f_veri be false; //Model verification pending;
5:        Let T_start = 0, T_end = 0, T_fin = 0;
6:        Let L = 0; // Initializes the user's potential step length;
7:        Let S = ""; // Initializes an empty string to store the user's gait segment;
8:        while an accelerometer instance has been received {
9:            Let the instance be A_i and convert A_i into R_i using Equation 1;
10:           if f_dect = false {
11:               Keep R_i;
12:               if i = 46 { Set T_start as the index of the minimum value in [R_1, R_2, ..., R_46]; }
13:               if i = T_start + 80 {
14:                   Set T_end as the index of the minimum value in [R_T_start+30, ..., R_T_start+80];
15:                   L = T_end − T_start;
16:               if L ≠ 0 and i = T_start + 8L{
17:                   Set T_fin as the index of the minimum value in [R_T_start+7L, ..., R_T_start+8L];
18:                   S = R_T_start, R_T_start+1, ..., R_T_fin;
19:                   Invoke BaseGen by sending S and L;
20:                   Set f_dect to be true when receiving a completion notification from BaseGen;
21:                   Jump to line 28;}}
22:           else if f_dect = true & f_veri = false {
23:               if i < T_start + L ∗ (8 + F){
24:                   Invoke OLAD by sending R_i;
25:                   if the output of OLAD indicates an anomaly { //i.e., the verification fails.
26:                       Clear and reset everything and inform the user to restart GAD from line 1; }
27:                   else{ Set f_veri to be true; break the while loop; }}}
28:           i = i + 1; }}
29:   if receiving a request for anomaly detection {
30:       Let E = [ ]; // Initializes an empty list for anomaly detection
31:       Let f_min be false; //The first minimum RAM value for anomaly detection not found;
32:       while an accelerometer instance has been received {
33:           Convert the instance into a RAM value using Equation 1;
34:           if the length of E < 46 { Append the RAM value to list E; }
35:           else {
36:               if f_min = true { Invoke OLAD by sending the RAM value;}
37:               else {
38:                   Invoke OLAD by sending the minimum value and remaining values in E;
39:                   Set f_min to be true ; }
40:           Notify the user of the anomaly output from OLAD; }}}
```

**Fig. 1.** The pseudo code of GAD.

Fig. 1 illustrates the pseudo code of GAD. Once initiated, GAD starts collecting accelerometer instances from the user. Let $A_i$ be the accelerometer instance at index $i$, where $i$ starts from 1. Unlike other methods that generate or rely on multiple features extracted from accelerometer data, GAD simplifies its process by converting each observed instance into $R_i$, as referred to as the resultant acceleration magnitude (RAM) hereafter, based on Equation 1 used in [4].

$$R_i = \sqrt{A_{x,i}^2 + A_{y,i}^2 + A_{z,i}^2} \qquad (1)$$

where $A_{x,i}$, $A_{y,i}$, and $A_{z,i}$ represent the x-axis, y-axis, and z-axis values of $A_i$, respectively. To capture the user's walking pattern, GAD collects a short gait segment using lines 11 to 18 of Fig. 1. It searches for the minimum value within the first 46 RAM values ($R_1$, $R_2$, ..., $R_{46}$) and designates the corresponding index as $T_{start}$. Subsequently, it identifies another minimum value within the next 30 to 80 RAM values, and marks the corresponding index as $T_{end}$. Please refer to Fig. 2 for a visual illustration. The difference between $T_{start}$ and $T_{end}$, denoted by $L$, suggests the user's potential step length. This method is called *personalized*. Afterward, GAD continues to collect data until another $7 * L$ RAM values have been generated, which suggests that the user may have taken another 7 steps. Upon locating the minimum value within the last $L$ RAM values and labeling its index as $T_{fin}$, the sequence of RAM values from $T_{start}$ to $T_{fin}$, as illustrated in Fig. 2, is considered as the user's gait segment, denoted by $S$. This design ensures that $S$ exhibits a recurrent pattern encompassing 8 gait cycles. The selection of the values 46, 30, and 80 is based on our prior experience with gait analysis.

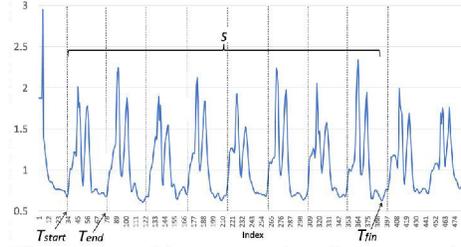

**Fig. 2.** Illustration of how a user's gait segment is derived.

Once $S$ is derived, it is immediately input into BaseGen, a component designed to build an anomaly detector for the user. This component will be further introduced in the following subsection. Upon receiving a completion notification from BaseGen, which indicates that an anomaly detector has been generated for the user, GAD proceeds to validate the detector using the user's subsequent $F$ walking steps, where $F$ represents a small integer. This process is referred to as *Model verification*, and it is done by sending each RAM value of the $F$ walking steps to OLAD for anomaly detection (see line 24). Note that the details of OLAD will be introduced later. If OLAD returns an anomaly, the verification is considered failed. In this case, the user will be instructed to restart GAD, which involves capturing a new gait segment from the user and training an anomaly detector from scratch. This design provides users with the flexibility to validate their anomaly detectors before employing them for future detection.

Conversely, if no anomaly is found, the verification is considered successful, and GAD then begins its anomaly detection operation at the user's request (see lines 29 and 40). Similar to line 12 of Fig. 1, GAD identifies the first minimum value among the subsequent 46 RAM values because it might signify the beginning of the user's walking step. It then begins sending that minimum value and every subsequent RAM value to OLAD for anomaly detection. It also reports back to the user whenever it receives an anomaly notification from OLAD.

### 3.1   Base Model Generator (BaseGen)

Inspired by SALAD [11], a self-adaptive anomaly detection approach for recurrent time series, and RePAD2 [10], a lightweight anomaly detection approach for nonrecurrent time series, BaseGen aims to construct an anomaly detector for the user. As shown in Fig. 3, BaseGen comprises four subcomponents arranged into two partially parallel processes. Firstly, Converter 1 (CV1) is paired with Detection Model Generator 1 (DMGen1), and secondly, Converter 2 (CV2) is paired with Detection Model Generator 2 (DMGen2).

Upon receiving the gait segment $S$ and $L$ from GAD, CV1 transforms $S$ into a series of AARE values, which are subsequently utilized by DMGen1 to train the first anomaly detection model. Concurrently, the same AARE values are also input into CV2, which further transforms them into another series of AARE values. The resulting AARE values are then utilized by DMGen2 to train the second anomaly detection model. This dual-model design aims to enhance anomaly detection accuracy.

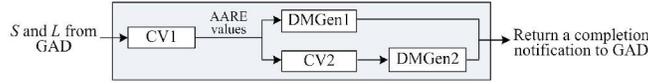

**Fig. 3.** The structure of BaseGen.

The detailed process is as follows: CV1 converts $S$ into an AARE series based on the conversion algorithm of SALAD [11]. This conversion aims to simplify the gait segment into a less complex AARE series, potentially facilitating more efficient anomaly detection. To illustrate the functioning of CV1, its process is visualized in Fig. 4. To simplify the explanation of this process, we renumber the indexes of all the RAM values in the gait segment $S$, namely, $T_{start}$, $T_{start}+1$, ..., $T_{fin}$, as 1, 2, ..., and $T_{fin} - T_{start}+1$, respectively. A window comprising the first $L$ values of this gait segment is used to train an LSTM-based prediction model, which then predicts the $(L + 1)^{\text{th}}$ value of the gait segment, denoted by $\widehat{R}_{L+1}$. After the prediction, the window shifts forward by one value. Hence, the window comprising the second value to $L + 1$ value of the gait segment is then utilized to retrain the prediction model for predicting the $(L + 2)^{\text{th}}$ value, denoted by $\widehat{R}_{L+2}$, and the same process continues.

To calculate an AARE value, CV1 requires the $L$ most recent pairs of actual RAM values and the corresponding predicted values (see Equation 2). Hence, it can calculate and output the first AARE value when index $i$ reaches $2L$, and this value is denoted by $AARE_{2L}$ (as shown in Fig. 4). Similarly, the second AARE value can be output at $2L + 1$, denoted by $AARE_{2L+1}$, and so forth.

$$AARE_i = \frac{1}{L} \cdot \sum_{g=i-L+1}^{i} \frac{\mid R_g - \widehat{R_g} \mid}{R_g}, i \geq 2L \qquad (2)$$

In Equation 2, $R_g$ represents the observed RAM value at index $g$ in $S$, and $\widehat{R_g}$ represents the predicted RAM value at the same index where $g$ ranges from $i - L + 1$ to $i$. Recall that the initial value of $i$ is 1. For instance, if $L$ is set to 46, the first AARE value is computed and output at index 92, using the pairs from index 47 to 92. Subsequently, the second AARE value is computed and output at index 93, using the pairs from index 48 to 93, and this pattern continues accordingly.

In order to maintain the effectiveness of the prediction model in predicting RAM values, CV1 calculates and updates a threshold, denoted by *thd*, using

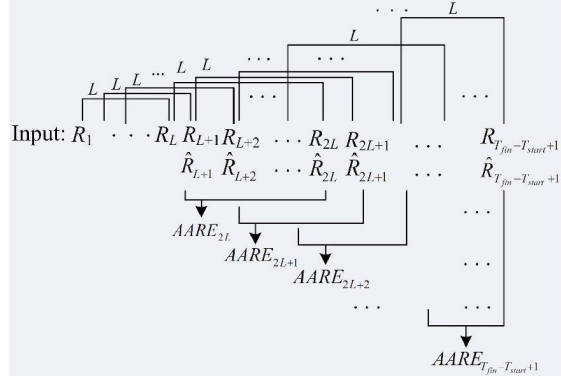

**Fig. 4.** Visualization of the prediction process of CV1.

Equation 3 following the Three-Sigma Rule [6]. It occurs whenever a new AARE value is derived and the total number of AARE values is not less than three.

$$thd = M_{AARE} + 3 \cdot \sigma \tag{3}$$

In Equation 3, $M_{AARE}$ represents the average of all previous AARE values, and $\sigma$ is the corresponding standard deviation. If the AARE value falls within $thd$, the prediction model is considered effective and is retained. However, if the AARE value exceeds $thd$, indicating that the current model is inadequate, CV1 retrains the model using the latest $L$ values of $S$.

Now, let us talk about how DMGen1 operates. To train the first anomaly detection model for the user, DMGen1 employs an algorithm similar to the one utilized by CV1. The key differences lie in the input, the window size, and the output. For DMGen1, the input is the AARE series output by CV1, its windows size is set to three, and its output is a notification about model completion. Similar to the design of RePAD2 [10], DMGen1 always utilizes the three latest input values to train and retrain its anomaly detection model. It calculates the AARE of the current detection model and updates its detection threshold using all previously derived AARE values (similar to Equation 3). Whenever the current detection model's AARE exceeds the current threshold, the model is replaced with a new one, trained using the latest three input values from CV1.

It is important to note that CV2 adopts a design similar to CV1 for LSTM model training, AARE computation, and detection threshold calculation, with the key difference being that CV2's input is the output from CV1. Similarly, DMGen2 adopts a design similar to DMGen1, but it utilizes the output of CV2 as its input to train the second anomaly detection model. Due to page limits, the detailed processes of CV2 and DMGen2 will not be repeated.

When both DMGen1 and DMGen2 complete their detection model training with their respective inputs, the anomaly detector generation for the user is considered completed. GAD is immediately informed about the completion, and it notifies the user. Note that all the knowledge acquired by each component of BaseGen will be retained and utilized by OLAD. This includes the latest prediction models, the latest detection models, the latest thresholds, the latest windows of their inputs and outputs, etc.

### 3.2   Online Anomaly Detection (OLAD)

OLAD has the same structure as BaseGen, as shown in Fig. 5. It comprises four subcomponents arranged in two partially parallel processes, with subcomponent Transformer 1 (TS1) paired with Anomaly Detection 1 (AD1), and Transformer 2 (TS2) paired with Anomaly Detection 2 (AD2). Upon receiving a RAM value, denoted by $R_j$, from GAD, OLAD passes it to TS1, which converts the value into an AARE value. This AARE value is then immediately sent to AD1 for anomaly detection. Simultaneously, the same AARE value is also passed to TS2 for further dimensionality reduction, and the resulting output is forwarded to AD2 for further anomaly detection. The introduction of AD2 is designed to detect those anomalies that AD1 is unable to detect.

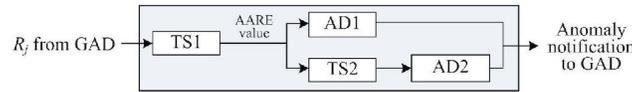

**Fig. 5.** The structure of OLAD.

It is worth mentioning that TS1, TS2, AD1, and AD2 in OLAD function similarly to CV1, CV2, DMGen1, and DMGen2 in BaseGen, respectively. However, they have the advantage of not starting from scratch because they inherit all the knowledge from BaseGen. As illustrated in Fig. 6, TS1 inherits the latest prediction model from CV1, enabling it to directly convert $R_j$ into an AARE value. TS1 only retrains the current prediction model when the model becomes inadequate for predicting RAM values, i.e., when the current AARE value exceeds the current detection threshold.

Similarly, AD1 inherits the latest anomaly detection model from DMGen1 for anomaly detection. Specifically, AD1 uses the detection model to predict the next input value, calculates the model's AARE, and updates its detection threshold based on all historical AARE values derived by DMGen1 and itself. If the current model's AARE exceeds the current threshold, AD1 retrains the model using the latest three input values. If the retrained model's AARE falls within the threshold, indicating a minor change in the user's walking pattern, nothing is reported. Conversely, if the AARE exceeds the threshold, the RAM value (i.e., $R_j$) is considered anomalous, and it is immediately reported to GAD.

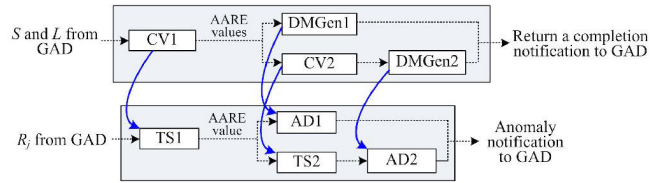

**Fig. 6.** Knowledge inheritance from the subcomponents of BaseGen to the subcomponents of OLAD.

Following the same principle, TS2 inherits the latest prediction model from CV2 and functions similarly to convert its input into an AARE value. Likewise, AD2 inherits the latest anomaly detection model from DMGen2 and functions similarly to determine whether $R_j$ is anomalous. If $R_j$ is found anomalous, it is immediately reported to GAD.

## 4    Evaluation and Results

To evaluate the detection performance of GAD, we utilized a dataset provided by the open-source OU-ISIR biometric database [13]. The dataset was collected manually using a IMU sensor positioned at the center back waist of 495 subjects. Each subject walked a predetermined path, and their walking sequences were collected, referred to as "walk1" in the dataset. The sensor is equipped with both a three-dimensional accelerometer and a three-dimensional gyroscope; however, we utilized the accelerometer data because its use is more common compare to the gyroscope, as mentioned in [7]. Utilizing this open-source dataset allows GAD to be easily compared with other approaches developed by researchers in the future.

We evaluated the detection performance of GAD by designing two scenarios: 1) where the step length parameter $L$ is tailored to each subject, i.e., the original design of GAD, and 2) where $L$ is consistently set to a fixed value, for example, 46 in our experiment. The first scenario is referred as **Personalized-GAD**, and the second scenario is referred as **Uniform-GAD** hereafter. The purpose is to determine the effectiveness of GAD by assessing whether a customized $L$ for each individual enhances accuracy, or if a uniform $L$ value offers a generalized and equally effective solution across different subjects.

It is important to note that existing gait anomaly detection approaches predominantly rely on offline data preprocessing and model training. In these approaches, gait data is collected and processed as a whole, and models undergo pre-training. Once the models are trained, they are deployed for detection without further updates. In contrast, GAD's online operational model stands out by requiring minimal preprocessing of gait data and enabling real-time, online training of anomaly detection models without the need for parameter configuration. Due to this fundamental difference in methodology, a direct comparison between these approaches and GAD is unfair as it would not accurately reflect the strengths and limitations of each. The most relevant approach to GAD is that proposed by Cola et al. [2]; however, as mentioned earlier in our related work section, this approach requires to extract 11 features, determine 2 parameters in advance, and requires several days to collect training data for each individual. All these requirements are unnecessary for GAD.

**Table 1.** Hyperparameter settings.

| Hyperparameter | Value |
| --- | --- |
| Hidden layer count | 1 for all models |
| Hidden unit count | 10 for all models |
| Learning rate | 0.0055 for all models trained by CV1, CV2, TS1 and TS2; 0.001 for all models trained by DMGen1, DMGen2, AD1, and AD2. |
| The number of epoch | Up to 100 for all models trained by CV1, CV2, TS1, and TS2; up to 50 for all models trained by DMGen1, DMGen2, AD1, and AD2. |
| Activation function | tanh for all models |
| Random seed | 140 for all models |

Table 1 lists all hyperparameters settings used by GAD in both scenarios, mostly following the configuration used by SALAD [11]. All LSTM models trained by BaseGen and OLAD for two evaluated scenarios feature a simple network structure, implemented in Deeplearning4j [3]. Each model consists of one hidden layer with 10 hidden units, aiming to keep each model simple and lightweight. To avoid model overfitting and underfitting, we utilized Early Stopping [5] to find the optimal epoch count. For models trained by CV1, CV2, TS1, and TS2, Early Stopping selected an epoch range of 1 to 100. For DMGen1, DMGen2, AD1, and AD2, which use fewer data points for model training, the range was narrowed to 1 to 50, in accordance with those used by SALAD. Note that we fixed all the aforementioned hyperparameter settings across all users without tailoring each one for individual users. While alternative settings might yield better results, conducting an exhaustive search for the optimal hyperparameter configuration for each user in an online manner is not feasible as those hyperparameters are needed to be pre-defined. Furthermore, to achieve a fair comparison, each experiment was separately conducted on a 2.6 GHz 6-Core Intel Core i7 MacBook running MacOS 10.15.4, with 16GB DDR4 SDRAM.

We replayed each subject's actual walking sequence by streaming it into GAD, which triggers BaseGen to generate an anomaly detector for each individual, followed by the corresponding model verification process. In both experiments, $F$ was set to 2, implying that the two subsequent steps of each individual were used for verification after their anomaly detectors had been generated. It is important to note that the choice of the value 2 was constrained by the length of the gait dataset utilized. When deploying GAD in a real-world scenario, it is recommended to slightly increase the value of $F$. As shown in Table 2, 135 out of the 495 subjects passed the model verification in the Personalized-GAD scenario, while 159 out of the 495 subjects passed it in the Uniform-GAD scenario. Nevertheless, the results suggest that tailoring $L$ to each individual's step length (i.e., the Personalized-GAD scenario) resulted in a stricter verification process, which led to a lower success rate compared with the Uniform-GAD scenario.

**Table 2.** Summary of Model Verification Results.

| Scenario | Number Passed | Number Failed | Total subjects |
|---|---|---|---|
| Personalized-GAD | 135 | 360 | 495 |
| Uniform-GAD | 159 | 336 | 495 |

With the previous results, we further evaluated the anomaly detection performance of GAD by concatenating X's gait segment and Y's gait segment, where X and Y are any two different subjects who passed the model verification process. Our purpose is to simulate a scenario where one person's identify is impersonated by another person. For instance, Fig. 7(a) shows the concatenation of the gait segment from subject T0_ID013843 followed by that of subject T0_ID310317. Conversely, Fig. 7(b) shows the concatenation starting from T0_ID310317's gait segment, followed by T0_ID013843's gait segment. Therefore, in Personalized-GAD, there are 18,090 (= $135^2 - 135$) segment combinations because 135 subjects passed their model verification in this scenario. On the other hand, in Uniform-GAD, there are 25,122 (= $159^2 - 159$) segment combinations because 159 subjects passed their model verification in this scenario.

Table 3 shows the anomaly detection performance of GAD in both scenarios. We can see that Personalized-GAD successfully identified 15,744 out of 18,090 segment combinations as originating from two different subjects. Hence, its detected anomaly ratio is 87.03%. On the other hand, in Uniform-GAD, only 19,869 out of 25,122 segment combinations were successfully identified as coming from two different subjects, resulting in a lower detected anomaly ratio, i.e., 79.09%. The above results suggest that when $L$ is tailored to each individual's step length, GAD generates an anomaly detector that is also tailored to each individual. This personalization allows for more effective detection of abnormal patterns. Therefore, it is recommended that GAD adopts the personalized approach to better adapt to individual variations and provide good anomaly detection.

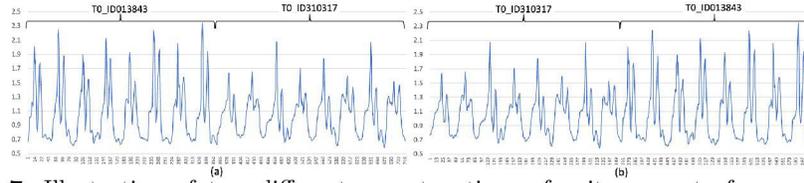

**Fig. 7.** Illustration of two different concatenations of gait segments from subjects T0_ID013843 and T0_ID310317.

**Table 3.** Anomaly Detection Performance of GAD in two scenarios.

| Scenario | Undetected Anomaly Ratio | Detected Anomaly Ratio |
|---|---|---|
| Personalized-GAD | 12.97% (= 2,346/18,090) | 87.03% (= 15,744/18,090) |
| Uniform-GAD | 20.91% (= 5,253/25,122) | 79.09% (= 19,869/25,122) |

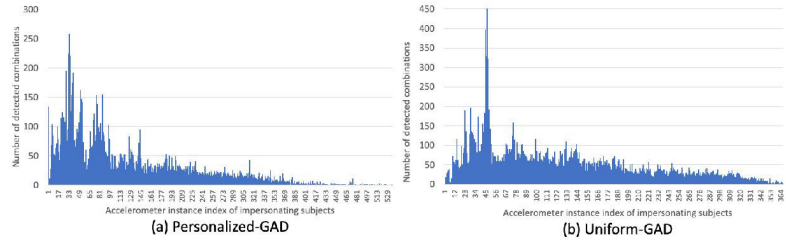

**Fig. 8.** Timing and quantity of successfully detected segment combinations.

To further understand GAD's efficiency in detecting anomalies, we illustrated the timing and quantity of successfully detected segment combinations in Fig. 8(a) and Fig. 8(b), which show the results for Personalized-GAD and Uniform-GAD, respectively. It is important to note that the term 'timing' in this context refers to the relative accelerometer instance index of subject Y (i.e., impersonating subject). Our aim is to know how promptly GAD can detect when one person is impersonated by another. From both figures, it is evident that both Personalized-GAD and Uniform-GAD are capable of early detection of impersonating subjects' gait segments as soon as these subjects begin walking. Specifically, Personalized-GAD successfully detected 11,102 out of the 15,744 impersonating subjects' gait segments upon analyzing their first 154 accelerometer instances. Similarly, Uniform-GAD successfully identified 13,304 out of the 19,869 impersonating subjects' gait segments upon processing their first 154 accelerometer instances. We can see that Uniform-GAD has higher efficiency than Personalized-GAD. Although Uniform-GAD is more efficient than Personalized-GAD, the latter provides a significantly higher anomaly detection ratio. Hence, it is still recommended that GAD adopts the personalized method.

## 5    Conclusions and Future Works

This paper presents GAD, a real-time gait anomaly detection approach that leverages an individual's three-dimensional accelerometer data. GAD stands out by utilizing dimensionality reduction and dual lightweight LSTM-based detection models to learn a user's gait pattern online using only a few steps of gait data. This eliminates the need for offline preprocessing, offline model training, offline model retraining, parameter setting, or threshold pre-determination. The online learning and adaptation features enable GAD to tolerate minor pattern changes and effectively identify significant deviations in a user's gait patterns. Furthermore, our exploration of both personalized and uniform methods for capturing a user's gait segment and our experiments using the open-source gait dataset demonstrate the superiority of the personalized approach in enhancing anomaly detection accuracy. GAD shows great potential in real-world applications, particularly as a crucial security measure in high-security areas. It can be used together with other authentication approaches in locations such as military bases, banks, nuclear facilities, data centers, ensuring that access to restricted areas is limited to authorized personnel.

In our future work, we plan to further enhance the detection performance of GAD by incorporating additional sensor data, e.g., gyroscope data, for greater accuracy. Furthermore, we intend to deploy GAD on wearable devices, such as smartphones for real-time and lightweight gait anomaly detection, which can be used for user authentication.

## Acknowledgement

The authors want to thank the anonymous reviewers for their reviews and valuable suggestions to this paper. This work has received funding from "the Research Council of Norway through the SFI Norwegian Centre for Cybersecurity in Critical Sectors (NORCICS) project no. 310105".